\documentclass[conference]{IEEEtran}
\IEEEoverridecommandlockouts

 \usepackage{color}
\usepackage{graphicx}
\usepackage{amsmath}
\usepackage{amssymb}

\newcommand{\be}{\begin{equation}}
\newcommand{\ee}{\end{equation}}
\newcommand{\ba}{\begin{array}}
\newcommand{\ea}{\end{array}}

\renewcommand{\thesubsubsection}{\thesubsection.\arabic{subsubsection}}

\title{Joint Hybrid Precoder and Combiner Design for mmWave Spatial Multiplexing Transmission}

\author{ \IEEEauthorblockN{Zihuan Wang$^{\dag}$, Ming Li$^{\dag}$,  Xiaowen Tian$^{\dag}$,  and Qian Liu$^{ \ddag}$
\vspace{-0.0 cm} }\\
\IEEEauthorblockA{$^{\dag}$School of Information and Communication Engineering   \\  Dalian University of Technology, Dalian, Liaoning 116024, China  \\
E-mail: \texttt{\{wangzihuan, tianxw\}@mail.dlut.edu.cn, mli@dlut.edu.cn}}

\IEEEauthorblockA{$^{\ddag}$  School of Computer Science and Technology \\  Dalian University of Technology, Dalian, Liaoning 116024, China \\ E-mail: \texttt{qianliu@dlut.edu.cn}} }

\begin{document}

\pagestyle{empty}

 \maketitle

\begin{abstract}
Millimeter-wave (mmWave) communications have been considered as a key technology for future 5G wireless networks because of the orders-of-magnitude wider bandwidth than current cellular bands. In this paper, we consider the problem of codebook-based joint analog-digital hybrid precoder and combiner design for spatial multiplexing transmission in a mmWave multiple-input multiple-output (MIMO) system. We propose to jointly select analog precoder and combiner pair for each data stream successively aiming at maximizing the channel gain while suppressing the interference between different data streams. After all analog precoder/combiner pairs have been determined, we can obtain the effective baseband channel. Then, the digital precoder and combiner are computed based on the obtained effective baseband channel to further mitigate the interference and maximize the sum-rate. Simulation results demonstrate that our proposed algorithm exhibits prominent advantages in combating interference between different data streams and offer satisfactory performance improvement compared to the existing codebook-based hybrid beamforming schemes.
\end{abstract}

\begin{keywords}
Millimeter-wave communication, hybrid precoder, multiple-input multiple-output (MIMO), antenna arrays, beamforming.
\end{keywords}

\maketitle

\section{Introduction}

The rapid proliferation of wireless devices has raised high demand of  increasingly high transmission data rate. Millimeter-wave (mmWave) wireless communications, operating in the frequency bands from 30-300 GHz, have been demonstrated to be an excellent  candidate to solve the spectrum congestion problem in recent experiments and are defining a new era of wireless communications because of the significantly large and unexploited mmWave frequency bands \cite{Pi CM 11}-\cite{Heath 16}. Economic and energy-efficient analog/digial hybrid precoding and combining transceiver architecture has been widely used in mmWave massive multiple-input multiple-output (MIMO) systems.

The hybrid precoding approach applies a large number of analog phase shifters to implement high-dimesional radio frequency (RF) precoders to compensate  the large path-loss at mmWave bands, and a small number of RF chains for low-dimensional digital precoders to provide the necessary flexibility to perform advanced multiplexing/multiuser techniques.
The major challenge in designing hybrid precoder is the practical constraints of the RF precoders, such as constant modulus, which is usually imposed by phase shifters. A popular solution to maximize the spectral efficiency of mmWave communications is to minimize the Euclidean distance between the hybrid precoder and the full-digital precoder. Thus, the hybrid precoder design is deemed as solving various matrix factorization problems with constant modulus constraints of the analog precoder \cite{Gao JSAC 16}-\cite{Rusu TWC}.

The existing hybrid precoding designs typically assume that the analog beamformers are implemented with  infinite resolution phase shifters.
However, the components for realizing accurate phase shifters could be very complicated and expensive.
Therefore, low-resolution phase shifters with discrete/quantized tunable phases are cost-effective and typically adopted in realistic systems.
Particularly, according to the special characteristic of a mmWave channel, more practical codebook-based hybrid precoder design has been widely used \cite{Ayach TWC 14}-\cite{Ayach Glob 13}, in which the columns of the analog precoder are selected from certain candidate vectors, such as array response vectors of the channel and discrete Fourier transform (DFT) beamformers which have constant modulus and discrete phases.

In the existing codebook-free and codebook-based algorithms mentioned above, the optimal hybrid precoder and combiner are individually designed to approximate, in the best Frobenius norm, the right and left singular vectors of the channel matrix, receptively.
While the separate design for hybrid precoder and combiner can provide satisfactory performance in terms of spectral efficiency, the orthogonality of resulting spatial multiplexing channel cannot be guaranteed.
Therefore, the conventional hybrid precoder and combiner designs may cause significant performance loss in realistic mmWave multiplexing systems.
This motivates us to reconsider the hybrid precoder and combiner design and find a better way for spatial multiplexing in mmWave MIMO communications.

In this paper, we consider the problem of codebook-based joint hybrid precoder and combiner design for spatial multiplexing transmission in mmWave MIMO systems. We propose to   jointly select the analog precoder and combiner pair for each data stream successively, which can maximize the channel gain as well as suppress the interference between different data streams. Then, the digital precoder and combiner are computed based on the obtained effective baseband channel to further mitigate the interference and maximize the sum-rate.   Simulation results demonstrate that our proposed algorithm exhibits prominent advantages in combating the interference between different data streams and offers satisfactory performance improvement compared to the existing codebook-based hybrid beamforming schemes.

The following notation is used throughout this paper. Boldface
lower-case letters indicate column vectors and boldface upper-case
letters indicate matrices; $\mathbb{C}$ denotes the set of all
complex numbers; $(\cdot)^T$ and $(\cdot)^H$  denote  the transpose and
transpose-conjugate operation, respectively; $\mathbf{A}(:,l)$ denotes the $l$-th column of matrix $\mathbf{A}$; $\mathbf{I}_L$ is the
$L\times L$ identity matrix; $\mathbb{E} \{ \cdot \}$ represents
statistical expectation. Finally, $| \cdot |$, $\| \cdot \|$, and $\| \cdot \|_F$ are the scalar magnitude, vector norm, and Frobenius norm, respectively.

\section{System Model and Problem Formulation}
\label{sc:system model}

\subsection{System Model}

\begin{figure}[!t]
\centering
%\vspace{-0.0 cm}
\includegraphics[width= 3.5 in]{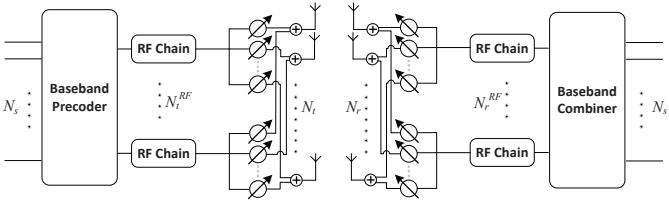}
\caption{A typical mmWave MIMO system with hybrid precoder and combiner.}\label{fig:system model}\vspace{-0.0 cm}
\end{figure}

We consider a single-user mmWave  MIMO multiplexing system with hybrid precoder and combiner, as illustrated in Fig. \ref{fig:system model}. The transmitter employs $N_t$ antennas and $N^{RF}_t$ RF chains to simultaneously transmit $N_s$ data streams to the receiver which is equipped with $N_r$ antennas and $N^{RF}_r$ RF chains. To ensure the efficiency of the communication with  limited number of RF chains, the number of
data streams is constrained as $N_s = N_t^{RF} = N_r^{RF}$, while the results can be applied to the general cases.

The transmitted symbols are first processed by a baseband
precoder $\mathbf{F}_{BB} \in \mathbb{C}^{N_t^{RF}\times N_s}$, then up-converted to the RF domain via $N_t^{RF}$ RF chains before being precoded with an analog precoder $\mathbf{F}_{RF}$ of dimension $ N_t \times N_t^{RF}$. While the baseband precoder $\mathbf{F}_{BB}$ enables both amplitude and phase modifications, the analog precoder $\mathbf{F}_{RF}$ has a constant amplitude $\frac{1}{\sqrt{N_t}}$ for each element since it is implemented using analog phase shifters.

The discrete-time transmitted signal can be written as
\begin{equation}
\mathbf{x} = \sqrt{P}\mathbf{F}_{RF} \mathbf{F}_{BB} \mathbf{s}
\end{equation}
where $\mathbf{s}$ is the $N_s \times 1$ symbol vector such that $\mathbb{E}\{ \mathbf{s} \mathbf{s}^H \} = \frac{1}{N_s}\mathbf{I}_{N_s}$. $P$ represents
transmit power,
 and the total transmit power constraint is enforced
by normalizing $\mathbf{F}_{BB}$ such that $\|\mathbf{F}_{RF}\mathbf{F}_{BB}\|^2_F = N_s$.

For simplicity, we consider a narrowband block-fading
propagation channel, which yields
the received signal as
\begin{equation}
\mathbf{y} = \sqrt{P} \mathbf{H} \mathbf{F}_{RF}\mathbf{F}_{BB}\mathbf{s} + \mathbf{n} \label{eq:received signal 1}
\end{equation}
where $\mathbf{y}$ is the $N_r \times 1$ received vector, $\mathbf{H}$ is the $N_r \times N_t$ channel matrix, and  $\mathbf{n}\thicksim \mathcal{CN}(\mathbf{0},\sigma^2\mathbf{I}_{N_r})$ is the complex Gaussian noise vector corrupting the received signal.

The receiver uses its $ N_r^{RF}$ RF chains and
phase shifters to obtain the processed receive signal which has a form of
\begin{equation}
\mathbf{\widehat{s}} = \sqrt{P}\mathbf{W}^H_{BB}\mathbf{W}^H_{RF} \mathbf{H} \mathbf{F}_{RF}\mathbf{F}_{BB}\mathbf{s} + \mathbf{W}^H_{BB}\mathbf{W}^H_{RF}\mathbf{n} \label{eq:received signal 2}
\end{equation}
where $\mathbf{W}_{BB}$ is the $N_r^{RF} \times Ns$ digital baseband combiner, $\mathbf{W}_{RF}$ is the $N_r \times N_r^{RF}$ analog RF combiner.

\subsection{Millimeter-Wave MIMO Channel Model}
Due to high free-space pathloss and large tightly-packed antenna arrays, the mmWave propagation in a massive MIMO system is well characterized by a limited spatial selectivity or scattering model, e.g. the Saleh-Valenzuela model, which allows us to accurately capture the mathematical structure of mmWave channels \cite{Ayach TWC 14}. The matrix channel $\mathbf{H}$ is assumed to be a sum contribution of $N_{cl}$ scattering clusters, each of which provides $N_{ray}$ propagation paths to the channel matrix $\mathbf{H}$. Therefore, the discrete-time narrow-band mmWave channel $\mathbf{H}$ can be formulated as
\begin{equation}
\mathbf{H}=\sqrt{\frac{N_t N_r}{N_{c1}N_{ray}}} \sum_{i=1}^{N_{cl}} \sum_{l=1}^{N_{ray}} \alpha_{il} \mathbf{a}_{r}(\theta_{il}^{r})\mathbf{a}_{t}(\theta_{il}^{t})^H
\end{equation}
where $\alpha_{il}\thicksim \mathcal{CN}(0,\sigma_{\alpha,i}^2)$ is the complex gain of the $l$-th propagation path (ray) in the $i$-th scattering cluster and it yields independent identically distribution (i.i.d.). Let $\sigma_{\alpha,i}^2$ represent the average power of the $i$-th cluster, and the total power satisfies $\sum_{i=1}^{N_{cl}}\sigma_{\alpha,i}^2 =N_{c1}$. $\theta_{il}^{t}$ and $\theta_{il}^{r}$ are the angles of departure (AoD) and the angle of arrival (AoA), respectively, which are assumed to be Laplacian-distributed with a mean cluster angle $\theta_{i}^{t}$ and $\theta_{i}^{r}$  and an angle spread of $\sigma_{\theta_i^t}$ and $\sigma_{\theta_i^r}$, respectively. Finally, the array response vectors $\mathbf{a}_{r}(\theta)$ and $\mathbf{a}_{t}(\theta)$ are the antenna array
response vectors, which only depend on the antenna array structures. When the commonly used uniform linear arrays (ULAs) are considered, the receive antenna array response vector can be written as
\begin{equation}
\mathbf{a}_r(\theta)=\frac{1}{\sqrt{N_r}}[1, {e}^{j\frac{2\pi}{\lambda}d\sin(\theta)}, \ldots , {e}^{j(N_r-1)\frac{2\pi}{\lambda}d\sin(\theta)}]^T  \vspace{-0.1 cm}
\end{equation}
where $\lambda$ is the signal wavelength, and $d$ is the distance between antenna elements. The transmit array response vector  $\mathbf{a}_t(\theta)$ can be written in a similar fashion.

\subsection{Problem Formulation}

We consider the problem of codebook-based hybrid precoder and combiner design in a mmWave multiplexing system.
Specifically, let $\mathcal{F}$ and $\mathcal{W}$ denote the beamsteering codebooks for the analog precoder and combiner, respectively. If $B^{RF}_t$ ($B^{RF}_r$) bits are used to quantize the AoD (AoA), $\mathcal{F}$ and $\mathcal{W}$ will consist of all possible analog precoding and combining vectors, which can be presented as
\begin{eqnarray}
\mathcal{F} & = &\{ \mathbf{a}_t(2 \pi i /2^{B_t^{RF}}) :   i=1, \ldots,   2^{B_t^{RF}}\}, \\
  \mathcal{W} & = &\{ \mathbf{a}_r(2 \pi i /2^{B_r^{RF}}) :  i=1, \ldots,   2^{B_r^{RF}}\}.
\end{eqnarray}
The columns of analog precoding (combining) matrix $\mathbf{F}_{RF}$ ($\mathbf{W}_{RF}$) are picked from candidate vectors in $\mathcal{F}$ ($\mathcal{W}$), i.e. $\mathbf{F}_{RF}(:,l) \in \mathcal{F}, \forall l =1, \ldots, N_t^{RF}$,  $\mathbf{W}_{RF}(:,l) \in \mathcal{W}, \forall l =1, \ldots, N_r^{RF}$.

When Gaussian symbols are transmitted over the mmWave MIMO channel,
the achieved spectral efficiency is given by
\begin{eqnarray}
R \hspace{-0.2 cm} & = & \hspace{-0.2 cm} \mathrm{log}_2 \Bigg( \bigg|  \mathbf{I}_{N_s} + \frac{P}{N_s} \mathbf{R}_n^{-1}  \mathbf{W}^H_{BB}\mathbf{W}^H_{RF} \mathbf{H} \mathbf{F}_{RF}\mathbf{F}_{BB} \times   \nonumber \\ & & \hspace{2.2 cm}  \mathbf{F}_{BB}^H \mathbf{F}_{RF}^H \mathbf{H}^H \mathbf{W}_{RF}  \mathbf{W}_{BB} \vspace{1cm}   \bigg| \Bigg), \label{eq:spectral efficiency}
\end{eqnarray}
where $\mathbf{R}_n \triangleq \sigma_n^2 \mathbf{W}^H_{BB}\mathbf{W}^H_{RF}\mathbf{W}_{RF}\mathbf{W}_{BB}$ is the noise covariance matrix after combining.

While most existing hybrid precoder and combiner design algorithms aim to maximize spectral efficiency in ({\ref{eq:spectral efficiency}}), we should point out that it is actually a performance upper bound for a general MIMO system. Even though precoder and combiner designs by approximating the right and left singular vectors of the channel matrix can provide satisfatory spectral efficiency performance, they cannot guarantee the orthogonality of the resulting effective spatial multiplexing channel for multiple data streams transmission.
Therefore, a practical spatial multiplexing system needs a more reasonable performance metrics, such as the sum-rate of each data stream which is described as follows.

Given the received signal in (\ref{eq:received signal 2}), the signal-to-interference-plus-noise ratio (SINR) of the $k$-th data stream is formulated by
\begin{equation}
\gamma_k =\frac{\frac{P}{N_s} \mid \mathbf{W}(:,k)^H \mathbf{H} \mathbf{F}(:,k)  \mid^2}{\frac{P}{N_s} \sum\limits_{\substack{i=1,i\neq k}}^{N_s}\mid \mathbf{W}(:,k)^H\mathbf{H}\mathbf{F}(:,i) \mid^2+\sigma_n^2\|\mathbf{W}(:,k)\|^2}
\label{eq:sinr}
\end{equation}
where $\mathbf{F} \triangleq \mathbf{F}_{RF} \mathbf{F}_{BB}$ and $\mathbf{W} \triangleq \mathbf{W}_{RF} \mathbf{W}_{BB}$. The achievable sum-rate of the spatial multiplexing system is
\begin{equation}
R_{\mathrm{sum}}=\sum_{k=1}^{K}\log(1+\gamma_k). \label{eq:sum-rate}
\end{equation}

In this paper, we aim to jointly design the precoders $\mathbf{F}_{RF}$, $\mathbf{F}_{BB}$ and combiners $\mathbf{W}_{RF}$, $\mathbf{W}_{BB}$ to maximize the sum-rate of the mmWave multiplexing system, which can be formulated as
\begin{equation}
\begin{aligned}
%&\underset{\substack{\mathbf{F_{RF}}, \mathbf{f}_{{BB}_k}, \mathbf{w}_k}}
&\hspace{-0.6 cm}\left\{\mathbf{F}_{RF}^\star, \mathbf{F}_{BB}^\star, \mathbf{W}_{RF}^\star, \mathbf{W}_{BB}^\star  \right\}
=\arg\max\sum_{k=1}^{N_s} \log \left( 1+\gamma_k\right) \\
&\hspace{0.8 cm} \textrm{s. t.} ~~~~\mathbf{F}_{RF}(:,l) \in \mathcal{F}, \forall l =1, \ldots, N_t^{RF}, \\
&\hspace{1.8 cm}  \mathbf{W}_{RF}(:,l) \in \mathcal{W}, \forall l =1, \ldots, N_r^{RF}, \\
&\hspace{1.8 cm} \|\mathbf{F}_{RF}\mathbf{F}_{BB}\|^2_F = N_s.
\label{eq:objective_function}
\end{aligned}
\end{equation}

The optimization problem of (\ref{eq:objective_function}) is obviously a non-convex NP-hard problem. In the next Section, we turn to seek a sub-optimal joint hybrid precoder and combiner design to reduce the complexity while achieving a satisfactory performance.

\section{Proposed Joint Hybrid Precoder and Combiner Design}
\label{sec:Proposed Design Full Connect}

To implement an efficient multiplexing system and maximize the sum-rate in (\ref{eq:sum-rate}), we need to design the precoder and combiner which can enhance the channel gain of each data stream as well as suppress the interference from each other.
To this end, we propose to decompose the difficult optimization problem in (\ref{eq:objective_function}) into a series of suboptimal problems which are much easier to be solved.
In particular, by considering transmit/receive RF chain pairs
one by one, we successively select analog precoder and combiner to maximize the corresponding channel gain while suppressing the co-channel interference. Then, the baseband digital precoder and combiner are computed to further mitigate the interference and maximize the sum-rate.

For the first data stream channel (i.e. $k=1$), we attempt to find the optimal analog precoder and combiner pair $(\mathbf{f}_{{RF}_1}^{\star}, \mathbf{w}_{{RF}_1}^{\star})$ from codebooks in order to obtain the largest beamforming gain:
\begin{equation}
 \left\{\mathbf{f}_{{RF}_1}^{\star},\mathbf{w}_{{RF}_1}^{\star}  \right\}= \textrm{arg}\underset{\substack{\mathbf{w}_{RF}  \in \mathcal{W} \\ \mathbf{f}_{RF}  \in \mathcal{F}}} {\textrm{max}} | \mathbf{w}^H_{RF} \mathbf{H}\mathbf{f}_{RF} |
\label{eq:beam_sel}
\end{equation}
which can be easily solved by searching all candidate vectors in $\mathcal{F}$ and $\mathcal{W}$ with computational complexity $O(|\mathcal{F}| |\mathcal{W}|)$. Then, we assign $\mathbf{f}^\star_{RF_1} $ and $\mathbf{w}^\star_{RF_1} $  to the precoding and combining matrices \begin{equation} \mathbf{F}^\star_{RF}(:,1) = \mathbf{f}^\star_{RF_1}, \end{equation}
\begin{equation}  \mathbf{W}^\star_{RF}(:,1) = \mathbf{w}^\star_{RF_1} .\end{equation}

For the rest $K-1$ data streams, we attempt to successively select precoders and combiners to actively avoid the interference of the data streams whose precoders and combiners have been already determined.
In particular, the component of previous determined precoders and combiners should be removed from the other data streams' channels such that similar analog precoders and combiners will not be selected by the other data streams. To achieve this goal, we first initialize $\mathbf{\widetilde{H}}=\mathbf{H}$, which will be updated successively in each step of selecting the analog beamformer pair. Let $\mathbf{p}_1 \triangleq \mathbf{f}^\star_{{RF}_1}$ and  $\mathbf{q}_1 \triangleq \mathbf{w}^\star_{{RF}_1}$ be the components of the determined analog precoder and combinder for the first data stream, respectively.
Then, before choosing the next analog precoder and combiner pair,
the channel should be updated by eliminating such a series of orthogonal components of previous determined analog beamformer pairs. Particularly, before finding the second  (i.e. $k=2$) analog precoder and combiner pair, we need to update channel as
\begin{equation}
\mathbf{\widetilde{H}} = (\mathbf{I}_{N_r} - \mathbf{q}_1\mathbf{q}_1^H) \mathbf{\widetilde{H}} (\mathbf{I}_{N_t} - \mathbf{p}_1\mathbf{p}_1^H)  \end{equation}
and then execute searching precess as
\begin{equation}
\left\{\mathbf{f}_{{RF}_2}^{\star},\mathbf{w}_{{RF}_2}^{\star}  \right\}= \textrm{arg}\underset{\substack{\mathbf{w}_{RF}  \in \mathcal{W} \\ \mathbf{f}_{RF}  \in \mathcal{F}}} {\textrm{max}} | \mathbf{w}^H_{RF} \mathbf{\widetilde{H}}\mathbf{f}_{RF}|.
\end{equation}
The analog precoders and combiners for the rest data streams can be successively selected using the above procedure. Note that when $k>1$, the orthogonormal component $\mathbf{p}_k$ and $\mathbf{q}_k$ of the selected precoder and combiner $\mathbf{f}_{{RF}_k}^{\star}$ and $\mathbf{w}_{{RF}_k}^{\star}$ should be obtained by a Gram-Schmidt based procedure:
\begin{equation}
 \mathbf{p}_k = \mathbf{f}_{{RF}_k}^{\star} -\sum\limits_{i=1}^{k-1}\mathbf{p}^H_i  \mathbf{f}_{{RF}_k}^{\star} \mathbf{p}_i, \nonumber  \end{equation}
\begin{equation}  \mathbf{p}_k = \mathbf{p}_k/\|\mathbf{p}_k\|, k=2,\ldots,K;\end{equation}
\begin{equation}
\mathbf{q}_k =  \mathbf{w}_{{RF}_k}^{\star} - \sum\limits_{i=1}^{k-1}\mathbf{q}^H_i\mathbf{w}_{{RF}_k}^{\star}\mathbf{q}_i , \nonumber  \end{equation}
\begin{equation}   \mathbf{q}_k = \mathbf{q}_k/\|\mathbf{q}_k\|, k=2,\ldots,K.
\end{equation}

\begin{center}
\begin{table}[!t]  \vspace{0.0 cm}\caption{Joint Hybrid Precoder and Combiner Design}
\begin{center} \begin{small}
\begin{tabular}{l}
\hline \hline \\
\hspace{-0.2 cm} \textbf{Input:} \hspace{0.2cm} $\mathcal{F}$, $\mathcal{W}$, $\mathbf{H}$.\\
\hspace{-0.2 cm} \textbf{Output:} \hspace{0.2cm} $\mathbf{F}_{RF}^{\star}$, $\mathbf{F}_{BB}^{\star}$, $\mathbf{W}_{RF}^{\star}$, and $\mathbf{W}_{BB}^{\star}$.\\
\hspace{-0.2 cm} \textbf{Initialization} $\mathbf{\widetilde{H}} = \mathbf{H}$.\\
\hspace{-0.2 cm} \textbf{for} $k=1:N_s$ \\
\hspace{0.2 cm} $\left\{\mathbf{f}^\star_{RF_k},\mathbf{w}^\star_{RF_k}  \right\}= \textrm{arg}\underset{\substack{\mathbf{w}_{RF} \in \mathcal{W} \\ \mathbf{f}_{RF} \in \mathcal{F}}} {\textrm{max}} | \mathbf{w}^H_{RF} \mathbf{\widetilde{H}}\mathbf{f}_{RF}|;$\\
\hspace{0.2 cm} $\mathbf{F}^\star_{RF}(:,k) = \mathbf{f}^\star_{RF_k} $;\\
\hspace{0.2 cm} $\mathbf{W}^\star_{RF}(:,k) = \mathbf{w}^\star_{RF_k} $.\\
\hspace{0.2 cm} \textbf{if} $k=1$ \\
\hspace{0.5 cm} $\mathbf{p}_k=\mathbf{f}^\star_{RF_k}$, $\mathbf{q}_k=\mathbf{w}^\star_{RF_k}$. \\
\hspace{0.2 cm} \textbf{else}  \\
\hspace{0.5 cm} $\mathbf{p}_k= \mathbf{f}^\star_{RF_k} -\sum\limits_{i=1}^{k-1}\mathbf{p}^H_i  \mathbf{f}^\star_{RF_k} \mathbf{p}_i$, $\mathbf{p}_k=\mathbf{p}_k/\|\mathbf{p}_k\|$;\\
\hspace{0.5 cm} $\mathbf{q}_k=\mathbf{w}^\star_{RF_k} -\sum\limits_{i=1}^{k-1}\mathbf{q}^H_i \mathbf{w}^\star_{RF_k} \mathbf{q}_i$, $\mathbf{q}_k=\mathbf{q}_k/\|\mathbf{q}_k\|$.\\
\hspace{0.2 cm} \textbf{end if}\\
\hspace{0.2 cm} $\mathbf{\widetilde{H}} = (\mathbf{I}_{N_r} - \mathbf{q}_k\mathbf{q}_k^H) \mathbf{\widetilde{H}} (\mathbf{I}_{N_t} - \mathbf{p}_k\mathbf{p}_k^H)$.\\
\hspace{-0.2 cm} \textbf{end for}\\
\hspace{-0.2 cm} Obtain $\mathbf{F}^\star_{{BB}}$ and  $\mathbf{W}^\star_{{BB}}$ by (\ref{eq:effective_channel})-(\ref{eq:digital_normalize}).\\
\hline
\vspace{-0.5 cm}
\end{tabular}\label{tb:full connect}%\vspace{-0.6cm}
\end{small}
\end{center}
\end{table}
\end{center}

After all analog beamformer pairs have been determined, we can obtain the effective baseband channel as
\begin{equation}
\mathbf{H}_{eff} = (\mathbf{W}_{RF}^{\star})^H \mathbf{H} \mathbf{F}_{RF}^{\star}  \label{eq:effective_channel}
\end{equation}
where $\mathbf{F}_{RF}^{\star} \triangleq [\mathbf{f}^\star_{RF_1}, \ldots, \mathbf{f}^\star_{RF_{N_s}}] $ and $\mathbf{W}_{RF}^{\star} \triangleq [\mathbf{w}^\star_{RF_1}, \ldots, \mathbf{w}^\star_{RF_{N_s}}]$. For the baseband precoder and combiner design, we perform singular value decomposition (SVD)
\begin{equation}
\mathbf{H}_{eff} = \mathbf{U} \mathbf{\Sigma} \mathbf{V},
\end{equation}
where $ \mathbf{U}$ is an $N_r \times N_r$ unitary matrix, $\mathbf{\Sigma}$ is an $N_r \times N_t$  diagonal matrix of singular values arranged in decreasing order, and $\mathbf{V}$ is an $N_t \times N_t$ unitary matrix. Then, an SVD-based baseband digital precoder is employed to further suppress the interference and maximize the sum-rate:
\begin{eqnarray}
 \mathbf{F}^\star_{{BB}}  =  \mathbf{V}(:,1:N_s), \\
 \mathbf{W}^\star_{{BB}}  =  \mathbf{U}(:,1:N_s).
\end{eqnarray}
Finally, we normalize the baseband precoder $\mathbf{F}^\star_{{BB}}$ by
\begin{equation}
\mathbf{F}^\star_{{BB}}=\frac{\sqrt{N_s}\mathbf{F}^\star_{{BB}}}{\|\mathbf{F}^\star_{RF}\mathbf{F}^\star_{{BB}}\|_F}.
\label{eq:digital_normalize}
\end{equation}
This joint hybrid precoder and combiner design algorithm is summarized in Table \ref{tb:full connect}.

\section{Simulation Results}
\label{sc:Simulation}

In this section, we illustrate the simulation results of the proposed joint hybrid precoder and combiner design. Both transmitter and receiver are equipped with a $128$-antenna ULA and the antenna spacing is $d=\frac{\lambda}{2}$. The number of RF chains at the transmitter and receiver are $N_{t}^{RF} = N_{r}^{RF} = 4$, so is the number of data streams, $N_s = 4$.  The channel parameters are set as $N_{cl} = 10$ clusters, $N_{ray} = 10$ rays per cluster, and the
average power of the $i$-th cluster is $ \sigma^2_{\alpha,i} = c\frac{7}{10}^i$ where $c=(\sum_{i=1}^{N_{cl}}(\frac{7}{10})^i)^{-1}N_{cl}$. The azimuths of the AoAs/AoDs within a cluster are assumed to be Laplacian-distributed with an angle spread of $\sigma_{\theta_i^r}=\sigma_{\theta_i^t}=2.5^\circ$. The mean cluster AoDs are assumed to be uniformly distributed over $[0,2\pi]$, while the mean cluster AoAs are uniformly distributed over an arbitrary $\frac{\pi}{3}$ sector. Finally, we employ a codebook consisting of array response vectors with $64$ uniformly quantized angle resolutions.

\begin{figure}[!t]
\centering
  \includegraphics[width=3.45 in]{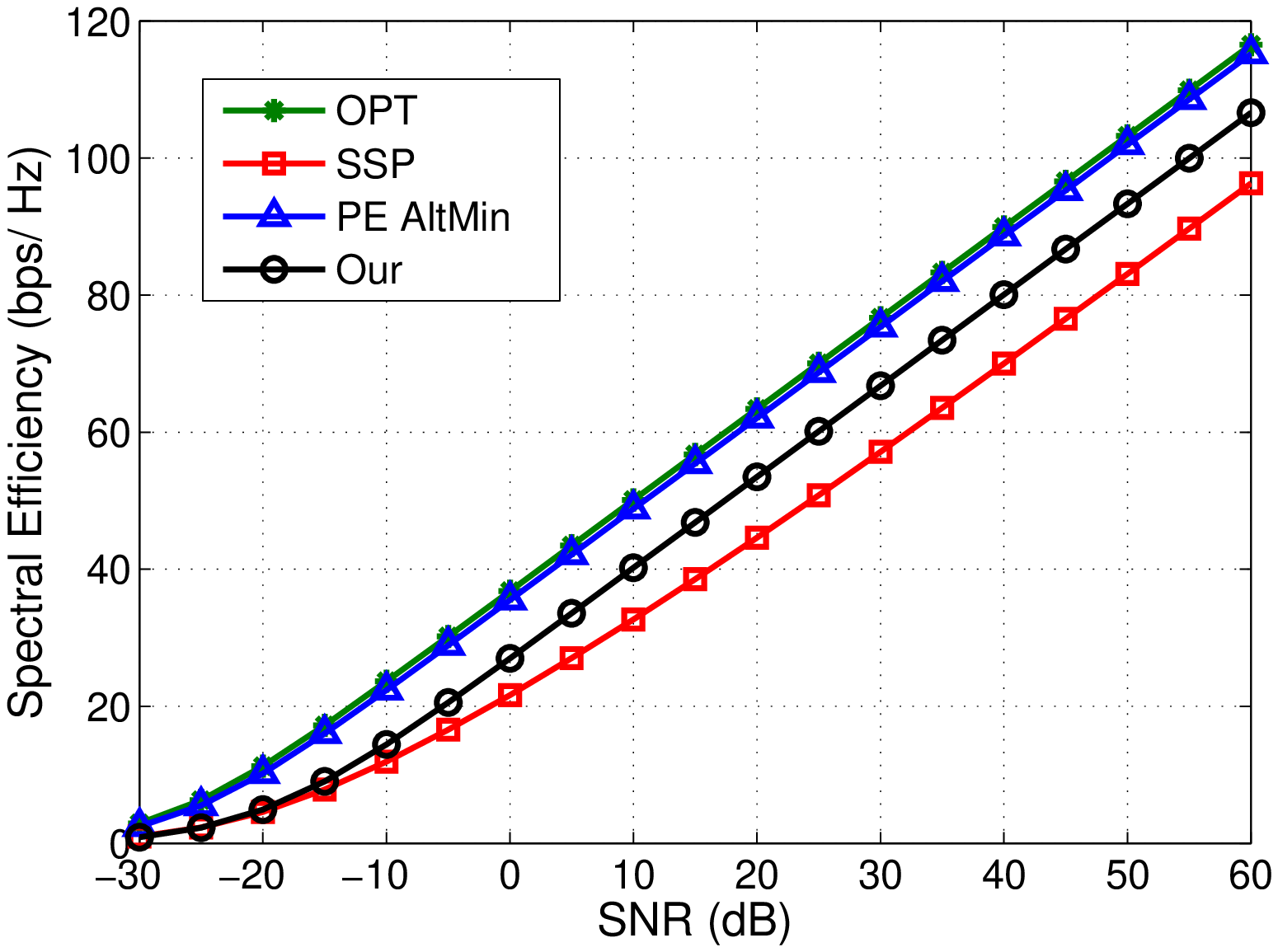}
  \vspace{-0.3 cm}
  \caption{Spectral efficiency versus SNR ($N_t= N_r=128$, $N_t^{RF}= N_r^{RF}=4$, $N_s=4$).}\label{fig:Full_C_N128_Ns4}%\vspace{-0.4 cm}

  \includegraphics[width=3.45 in]{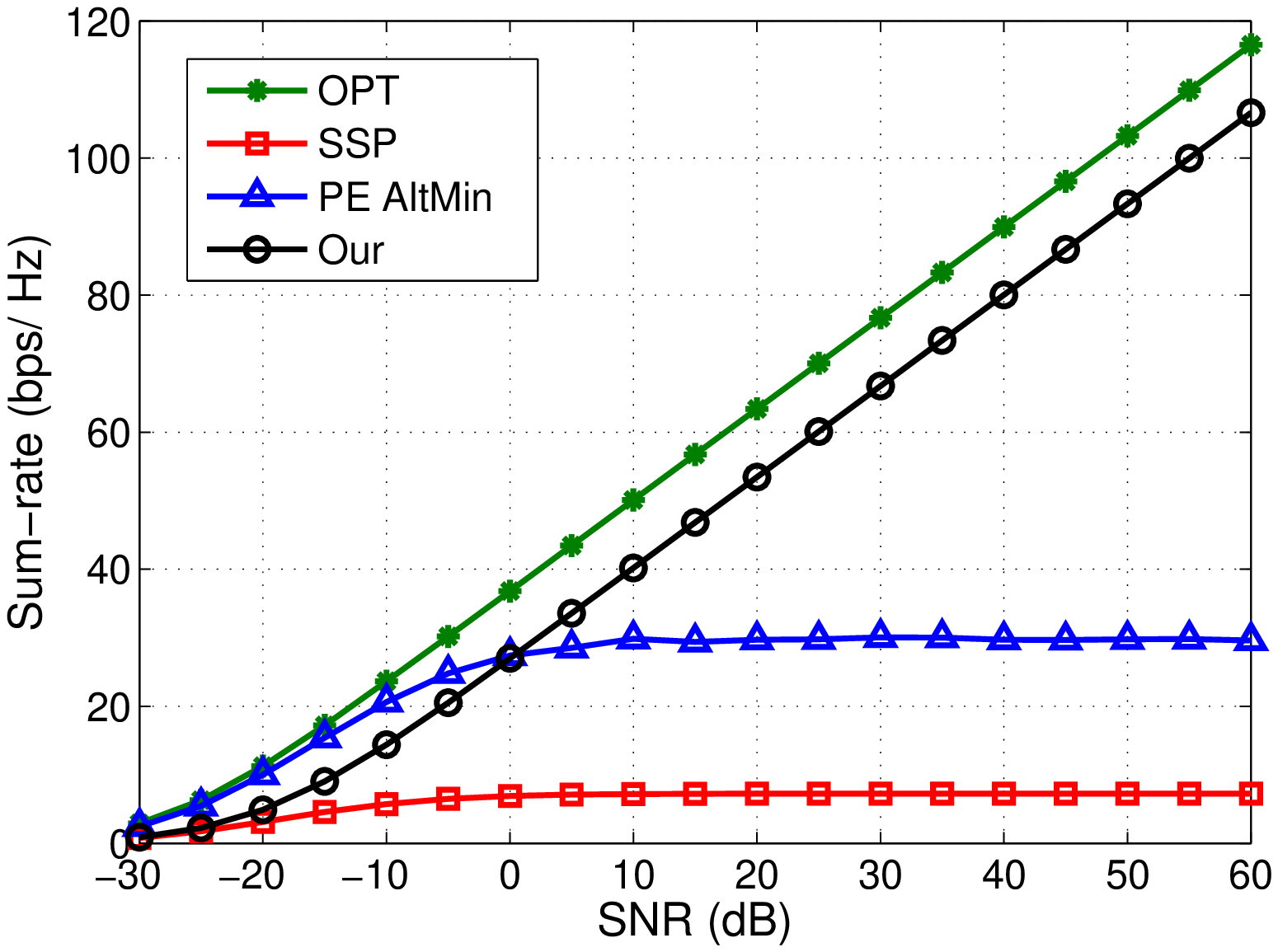}
  \vspace{-0.3 cm}
  \caption{Sum-rate versus SNR ($N_t= N_r=128$, $N_t^{RF}= N_r^{RF}=4$, $N_s=4$).}\label{fig:Full_R_N128_Ns4}\vspace{-0.3 cm}

\end{figure}

\begin{figure}[!t]
\centering
  \includegraphics[width=3.45 in]{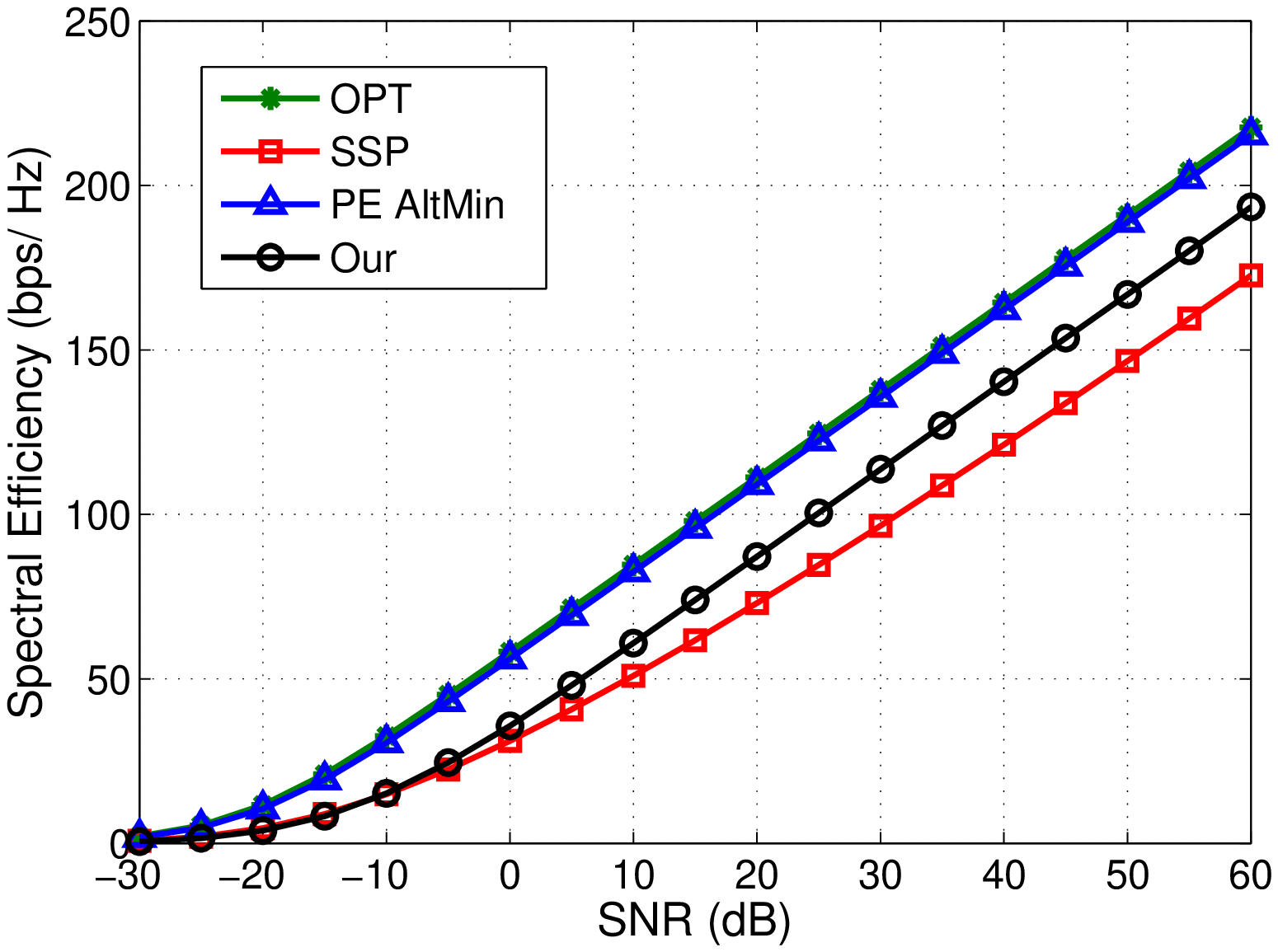}
  \vspace{-0.3 cm}
  \caption{Spectral efficiency versus SNR ($N_t= N_r=128$, $N_t^{RF}= N_r^{RF}=8$, $N_s=8$).}\label{fig:Full_C_N128_Ns8}%\vspace{-0.4 cm}

  \includegraphics[width=3.45 in]{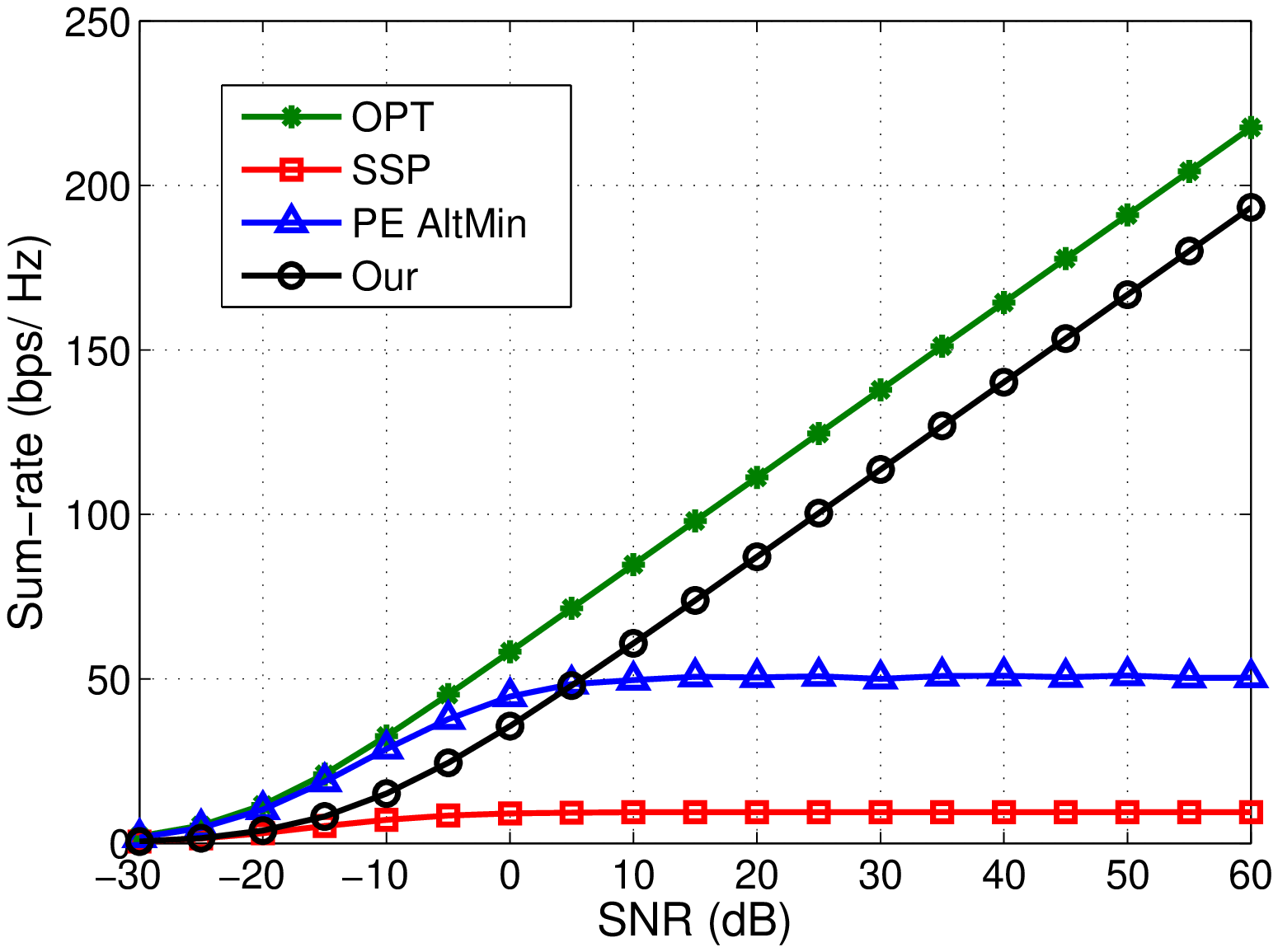}
  \vspace{-0.3 cm}
  \caption{Sum-rate versus SNR ($N_t= N_r=128$, $N_t^{RF}= N_r^{RF}=8$, $N_s=8$).}\label{fig:Full_R_N128_Ns8}
\end{figure}

\begin{figure}[!t]
\centering
  \includegraphics[width=3.4 in]{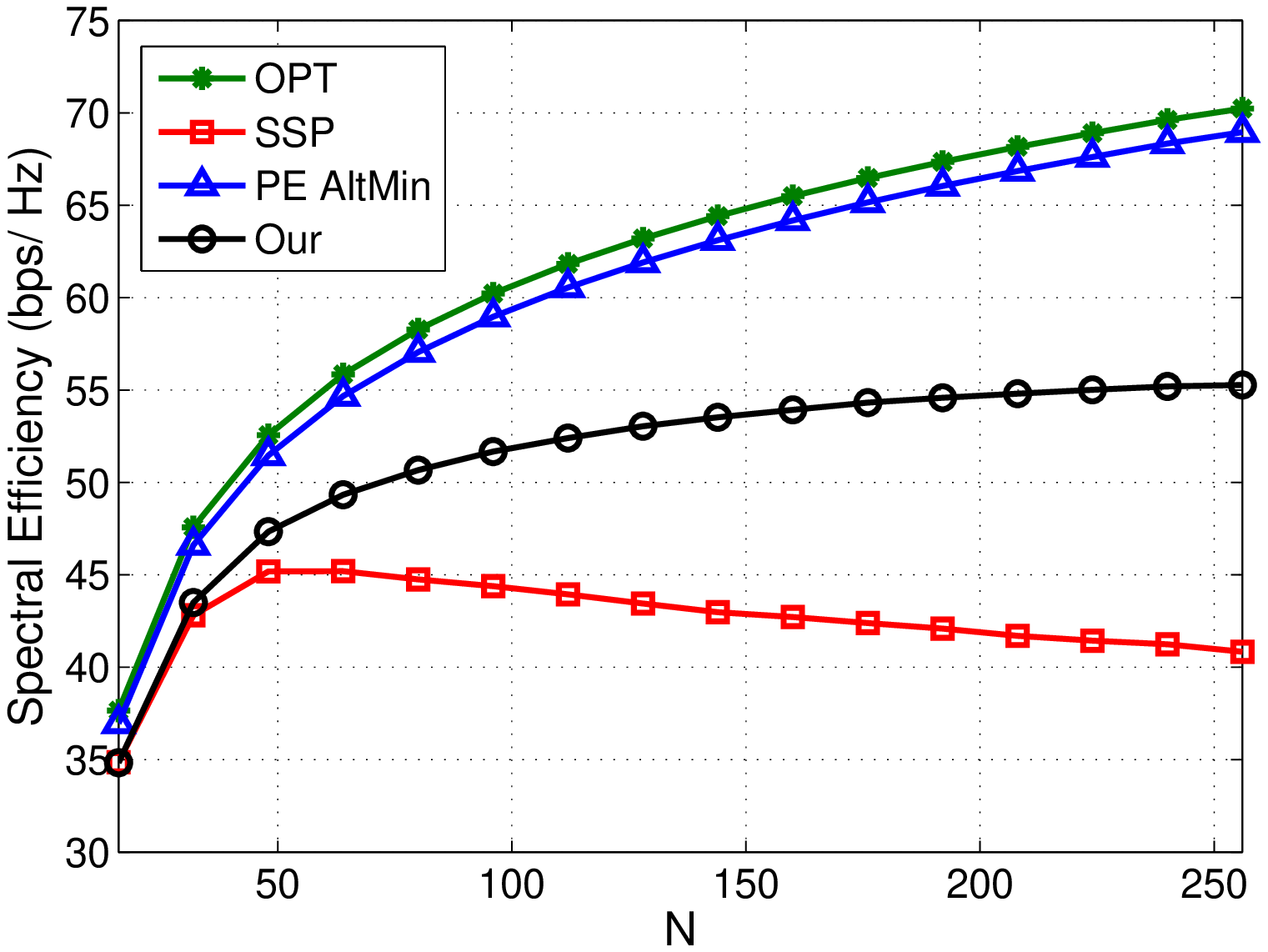}
  \vspace{-0.3 cm}
  \caption{Spectral efficiency versus the number of antennas (SNR=$20$dB, $N_t^{RF}= N_r^{RF}=4$, $N_s=4$).}\label{fig:Full_N_vs_se}

\centering
  \includegraphics[width=3.4 in]{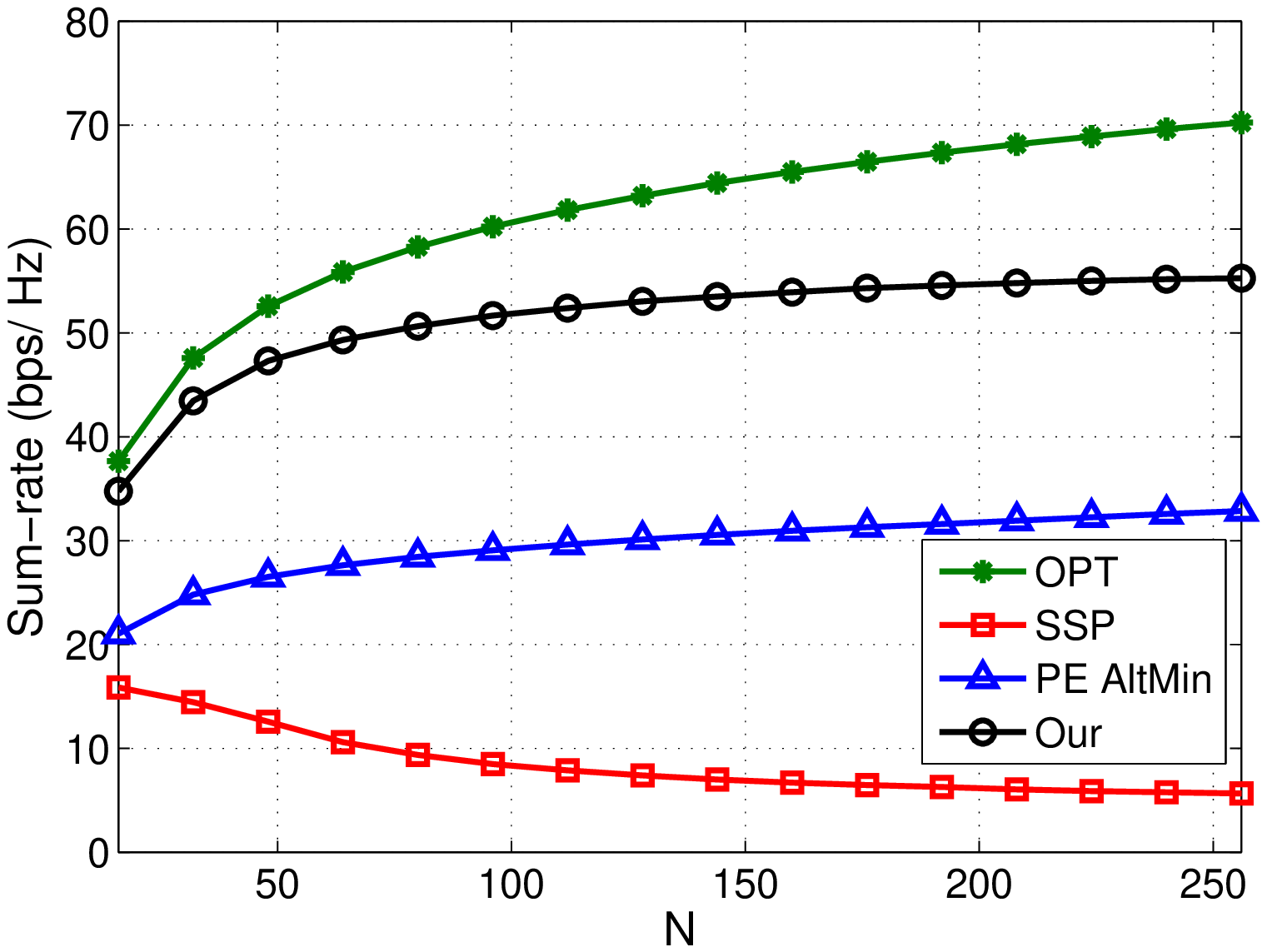}
  \vspace{-0.3 cm}
  \caption{Sum-rate versus the number of antennas (SNR=$20$dB, $N_t^{RF}= N_r^{RF}=4$, $N_s=4$).}\label{fig:Full_N_vs_rate}\vspace{-0.0 cm}

\end{figure}

\begin{figure}[!t]
\centering

\centering
  \includegraphics[width=3.4 in]{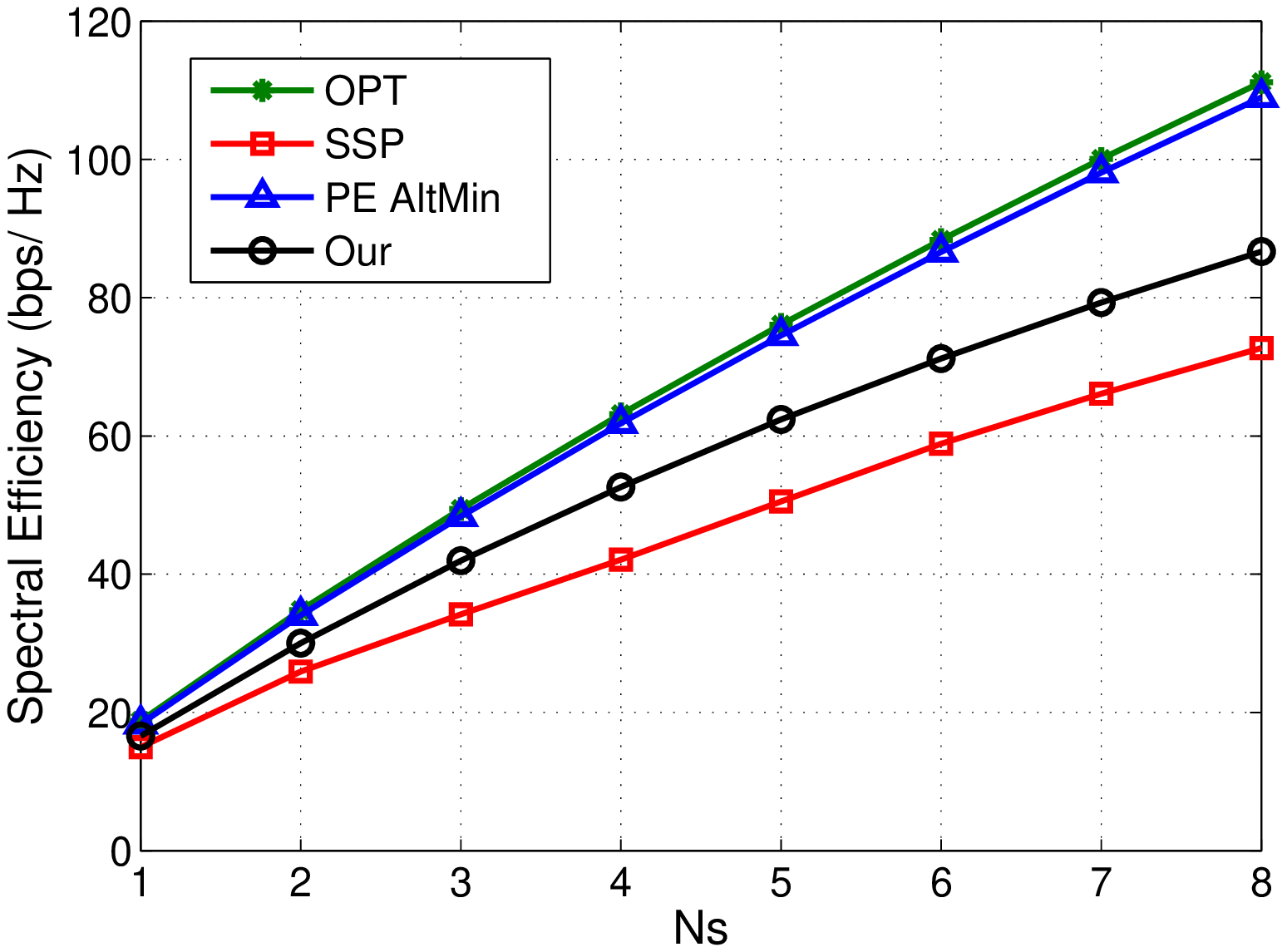}
  \vspace{-0.3 cm}
  \caption{Spectral efficiency versus $N_s$ ($N_t= N_r=128$, SNR=$20$dB).}\label{fig:Full_Ns_vs_se}

  \vspace{0.3 cm}

\centering
  \includegraphics[width=3.4 in]{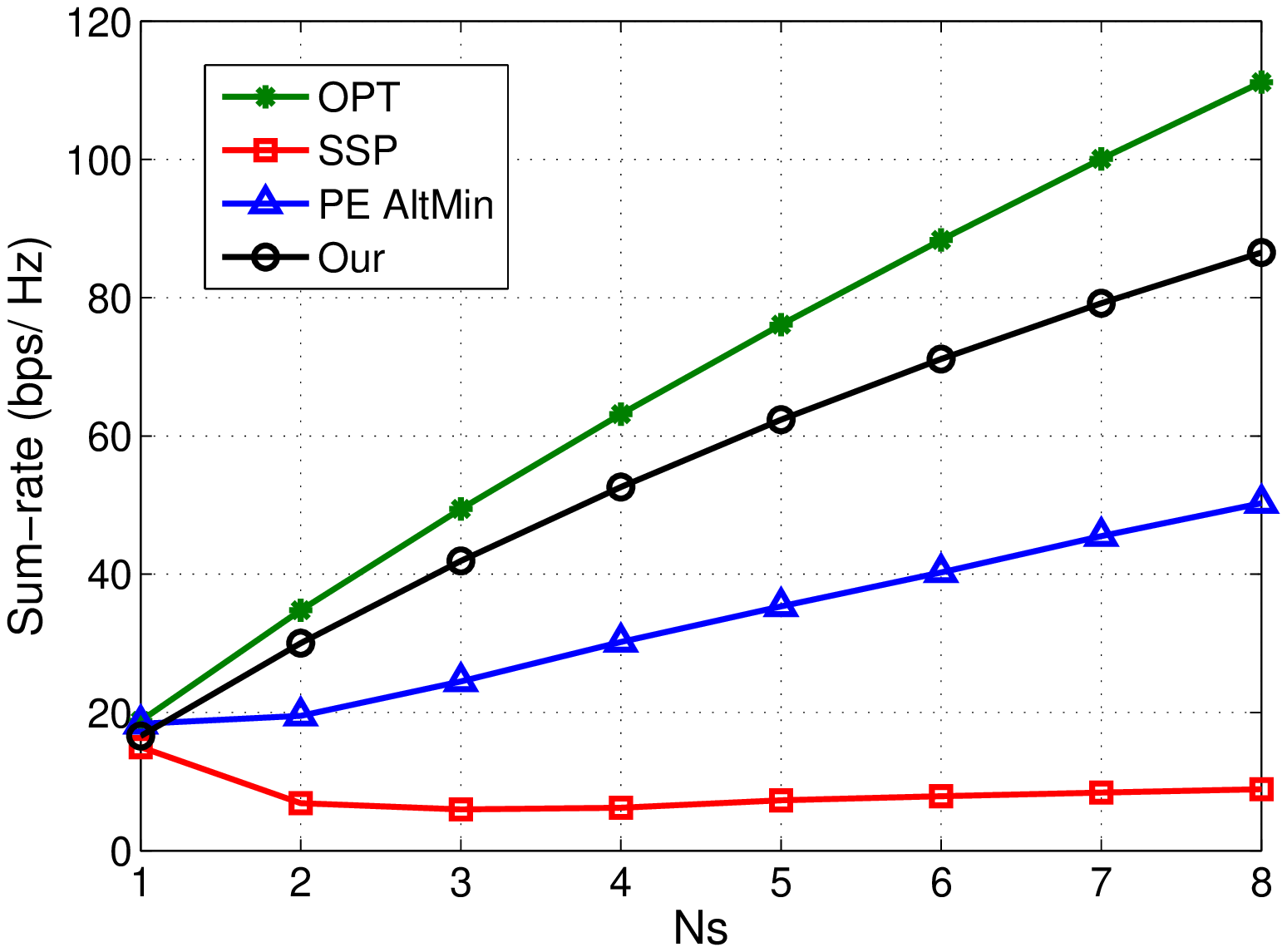}
  \vspace{-0.3 cm}
  \caption{Sum-rate versus $N_s$ ($N_t= N_r=128$, SNR=$20$dB).}\label{fig:Full_Ns_vs_rate} \vspace{0.3  cm}

\end{figure}

Fig. \ref{fig:Full_C_N128_Ns4} shows the spectral efficiency versus signal-to-noise-ratio (SNR) over $10^6$ channel realizations. For the comparison purpose, we also include two state-of-the-art algorithms: \textit{i}) Spatially Sparse Precoding (SSP) \cite{Ayach TWC 14}, which is a classic codebook-based hybrid precoding design; \textit{ii}) Alternating Minimization using Phase Extraction (PE AltMin) algorithm \cite{Yu JSAC 16}, which is a codebook-free hybrid precoding design. The optimal (OPT) full-digital beamforming scheme with the unconstrained SVD algorithm is also plotted as the performance benchmark. It can be observed that our proposed algorithm outperforms the codebook-based SSP algorithm. Note that the PE AltMin algorithm has a continuous phase on RF beamformer and can achieve extremely close performance to the optimal full-digital approach. Therefore, we just consider it as a reference for the codebook-free algorithms will not compare it with our proposed algorithm.
Fig. \ref{fig:Full_R_N128_Ns4} presents the sum-rate versus SNR with the same system settings as of Fig. \ref{fig:Full_C_N128_Ns4}. We can notice that the proposed joint hybrid precoder and combiner design has a significant performance advantage over the other two hybrid beamforming designs. This is because our joint analog precoder and combiner selection approach aims to mitigate the interference between different data streams.
We also consider a different setting that the number of RF chains and data streams are both set as $8$. In Figs. \ref{fig:Full_C_N128_Ns8} and \ref{fig:Full_R_N128_Ns8}, the spectral efficiency and sum-rate are presented, respectively. From these two figures, we can draw similar conclusions that the proposed algorithm can significantly suppress the inter-stream interference and take advantages of spatial multiplexing.

In Figs. \ref{fig:Full_N_vs_se} and \ref{fig:Full_N_vs_rate}, we turn to illustrate how the number of transmit and receive antennas affects the spectral efficiency and sum-rate performance. We assume $N_t=N_r=N$ which is varying from $16$ to $256$. The SNR is set at $20$dB and $N_t^{RF}=N_r^{RF}=N_s=4$. It can be observed from these two figures that the proposed algorithm has significant superiority compared with the SSP approach in the spectral efficiency as well as sum-rate performance. Interestingly, we also notice that the SSP scheme may exhibit severe performance degradation when the resolution of codebook is less than the number of antennas. This phonomania is clear shown in Fig. \ref{fig:Full_N_vs_se} with a turning point around $N=64$.

Figs. \ref{fig:Full_Ns_vs_se} and \ref{fig:Full_Ns_vs_rate} provide the performance of spectral efficiency and sum-rate versus the number of data streams $N_s$, respectively. The number of transmit and receive RF chains varies along with $N_s$. We can see that three approaches can achieve comparable performance in terms of spectral efficiency. However, only our proposed algorithm can maintain a satisfactory sum-rate achievement with the increase number of data streams. In addition, the strong interference between different data streams even causes a performance loss of SSP algorithm.

\section{Conclusions}
\label{sc:Conclusions}

This paper investigated the problem of codebook-based joint hybrid precoder and combiner design for spatial multiplexing transmission in mmWave MIMO systems. We proposed to jointly select analog precoder and combiner pair for each data stream successively aiming at maximizing the channel gain as well as suppressing the interference between different data streams.
Then, the digital precoder and combiner were computed based on the obtained effective baseband channel to further mitigate the interference and maximize the sum-rate. Simulation results demonstrated the performance improvement of our proposed algorithm compared to the existing codebook-based hybrid beamforming schemes.


\begin{thebibliography}{99}

%%% overview of mmWave

\bibitem{Pi CM 11} Z. Pi and F. Khan, ``An introduction to millimeter-wave mobile broadband systems,'' \textit{IEEE Commun. Mag.}, vol. 49, no. 6, pp. 101-107, June 2011.

\bibitem{Rappaport IA 13} T. Rappaport, S. Sun, R. Mayzus, H. Zhao, Y. Azar, K. Wang, G. N. Wong, J. K. Schulz, M. Samimi and F. Gutierrez ``Millimeter wave mobile communications for 5G cellular: It will work!'' \textit{IEEE Access}, vol. 1, pp. 335-349, 2013.


\bibitem{Heath 16} R. W. Heath Jr., N. Gonz\'{a}lez-Prelcic, S. Rangan, W. Roh, and A. M. Sayeed, ``An overview of signal processing techniques for millimeter wave MIMO systems,'' \textit{IEEE J. Sel. Topics Signal Process.}, vol. 10, no. 3, pp. 436-453, Apr. 2016.



%\bibitem{Roh CM 14} W. Roh, J.-Y. Seol, J. Park, B. Lee, J. Lee, Y. Kim, J. Cho, K. Cheun, and F. Aryanfar, ``Millimeter-wave beamforming as an enabling technology for 5G cellular communications: Theoretical feasibility and prototype results,'' \textit{IEEE Commun. Mag.}, vol. 52, no. 2, pp. 106-113, Feb. 2014.
%
%
%
%
%\bibitem{Andrews JSAC 14} J. G. Andrews, S. Buzzi, C. Wan, S. V. Hanly, A. Lozano, A. C. K. Soong, and J. C. Zhang, ``What will 5G be?'' \textit{IEEE J. Sel. Areas Commun.}, vol. 32, no. pp. 1065-1082, June 2014.

%
%\bibitem{Wei CM 14}  L. Wei, R. Q. Hu, Y. Qian, and G.Wu, ``Key elements to enable millimeter wave communications for 5G wireless systems,'' \textit{IEEE Commun. Mag.}, vol. 21, no. 6, pp. 136每143, Dec. 2014.

%%%% Channel modeling
%
%
%\bibitem{Samimi ICC 15}  M. K. Samimi and T. S. Rappaport, ``3-D statistical channel models for millimeter-wave outdoor mobile broadband communications,'' in \textit{Proc. IEEE Int. Conf. Commun. (ICC)}, London UK, June 2015, pp. 2430-2436.


%
%\bibitem{Rappaprot 15} T. S. Rappapport, G. R. MacCartney, M. K. Samimi Jr., and S. Sun, ``Wideband millimeter-wave propagation measurements and channel models for future wireless communication system design,'' \textit{IEEE Trans. Commun.}, vol. 63, no. 9, pp. 3029-3056, Sept. 2015.






%%% continious phase
\bibitem{Gao JSAC 16} X. Gao, L. Dai, S. Han, C.-L. I, and R. W. Heath Jr., ``Energy-efficient hybrid analog and digital precoding for mmWave MIMO systems with large antenna arrays,'' \textit{IEEE J. Sel. Areas Commun.}, vol. 34, no. 4, pp. 998-1009, April 2016.


\bibitem{Dai ICC 15} L. Dai, X. Gao, J. Quan,   S. Han, and C.-L. I, ``Near-optimal hybrid analog and digital precoding for downlink mmWave massive MIMO systems,'' in \textit{Proc. IEEE Int. Conf. Commun. (ICC)}, London UK, June 2015, pp. 1334-1339.



\bibitem{Yu JSAC 16}  X. Yu, J.-C. Shen, J. Zhang, and K. B. Letaief, ``Alternating minimization algorithms for hybrid precoding in millimeter wave MIMO systems,'' \textit{IEEE J. Sel. Topics Signal Process.}, vol. 10, no. 3, pp. 485每500, Apr. 2016.
%
%\bibitem{Han CM 15} S. Han, C.-L. I, Z. Xu, and C. Rowell, ``Large-scale antenna systems with hybrid analog and digital beamforming for millimeter wave 5G,'' \textit{IEEE Commun. Mag.}, vol. 53, no. 1, pp. 186-194, Jan. 2015.

%
%
%\bibitem{Bogale TWC 16}  T. E. Bogale, L. B. Le, A. Haghighat, and L. Vandendorpe, ``On the number of RF chains and phase shifters, and scheduling design with hybrid analog-digital beamforming,''  \textit{IEEE Trans. Wireless Commun.}, vol. 15, no. 5, pp. 3311-3326, May 2016.
%
%\bibitem{Rusu ICC 15} C. Rusu, R. M\'{e}ndez-Rial, N. Gonz\'{a}lez-Prelcic, and R. W. Heath Jr., ``Low complexity hybrid sparse precoding and combining in millimeter wave MIMO systems,'' in \textit{Proc. IEEE Int. Conf. Commun. (ICC)}, London, UK, June 2015, pp. 1340每1345.


\bibitem{Rusu TWC} C. Rusu, R. M\'{e}ndez-Rial, N. Gonz\'{a}lez-Prelcic, and Robert W. Heath Jr., ``Low Complexity Hybrid Precoding Strategies for Millimeter Wave Communication Systems,'' \textit{IEEE Trans. Wireless Commun.}, vol. 15, no. 12, pp. 8380-8393, Dec. 2016.



%%% continious phase + multiuser
%
%\bibitem{Liang WCL 14} L. Liang, W. Xu, and X. Dong, ``Low-complexity hybrid precoding in massive multiuser MIMO systems,'' \textit{IEEE Wireless Commun. Lett.}, vol. 3, no. 6, pp. 653-656, Dec. 2014.
%
%
%
%\bibitem{Li 14} X. Li, S. Zhou, Y. Yan, Z. Xiao, and J. Wang, ``Cluster information based user scheduling for multiuser MIMO systems,'' \textit{IEEE Mil. Commun. Conf. (MILCOM)}, Baltimore, MD, Oct. 2014, pp. 1580-1585.
%
%\bibitem{Nguyen ICC 16} D. H. N. Nguyen, L. B. Le, and T. Le-Ngoc, ``Hybrid MMSE precoding for mmWave multiuser MIMO systems,'' \textit{IEEE Int. Conf. on Commun. (ICC)}, Kuala Lumpur, Malaysia, May 2016.
%
%\bibitem{Li CL 16} J. Li, L. Xiao, X. Xu, and S. Zhou, ``Robust and low complexity hybrid beamforming for uplink multiuser mmWave MIMO systems,'' \textit{IEEE Commun. Lett.}, vol. 20, no. 6, pp. 1140-1143, June 2016.
%
%\bibitem{Lee TCOM} C.-S. Lee and W.-H. Chung, ``Hybrid RF-baseband precoding for cooperative multiuser massive MIMO systems with limited RF chains,'' \textit{IEEE Trans. Commun.}, to Appeal.
%
%\bibitem{Rajashekar TWC}  R. Rajashekar and L. Hanzo, ``Iterative matrix decomposition aided block diagonalization for mm-Wave multiuser MIMO systems,'' \textit{IEEE Trans. Wireless Commun.}, to Appeal.

%%% quantized phase + code beam


\bibitem{Ayach TWC 14} O. E. Ayach, S. Rajagopal, S. Abu-Surra, Z. Pi, and R. W. Heath Jr., ``Spatially sparse precoding in millimeter wave MIMO systems,'' \textit{IEEE Trans. Wireless Commun.}, vol. 13, no. 3, pp. 1499-1513, Mar. 2014.


\bibitem{Alkhateeb JSTSP 14} A. Alkhateeb, O. El Ayach, G. Leus, and R. W. Heath Jr., ``Channel estimation and hybrid precoding for millimeter wave cellular systems,'' \textit{IEEE Journal of Selected Topics in Signal Processing}, vol. 8, no. 5, pp. 831-846, Oct. 2014.


\bibitem{Alkhateeb TWC 15} A. Alkhateeb, G. Leus, and R. W. Heath Jr., ``Limited feedback hybrid precoding for multi-user millimeter wave systems,'' \textit{IEEE Trans. Wireless Commun.}, vol. 14, no. 11, pp. 6481-6494, Nov. 2015.


\bibitem{Alkhateeb TCOM 14} A. Alkhateeb and R. W. Heath Jr., ``Frequency selective hybrid precoding for limited feedback millimeter wave systems,'' \textit{IEEE Trans.
Commun.}, vol. 64, no. 5, pp. 1801-1818, May 2016.


\bibitem{Lee TSP 15}  Y. Lee, C.-H. Wang, and Y.-H. Huang, ``A hybrid RF/baseband precoding processor based on parallel-index-selection matrix-inversion-bypass simultaneous orthogonal matching pursuit for millimeter wave MIMO systems,'' \textit{IEEE Trans. Signal Process.}, vol. 63, no. 2, pp. 305-317, Jan. 2015.


\bibitem{Kim TVT 15} M. Kim and Y. Lee, ``MSE-based hybrid RF/baseband processing for millimeter wave communication systems in MIMO interference channels,'' \textit{IEEE Trans. Veh. Technol.}, vol. 64, no. 6, pp. 2714-2720, June 2015.

%
%\bibitem{Lee ICC 14} J. Lee and Y. Lee, ``AF relaying for millimeter wave communication systems with hybrid RF/baseband MIMO processing,'' in Proc. \textit{IEEE Int. Conf. Commun. (ICC)}, Sydney, Australia, June 2014, pp. 5838-5842.



\bibitem{Gao TVT 16} X. Gao, L. Dai, C. Yuen, and Z. Wang, ``Turbo-like beamforming based on Tabu search algorithm for millimeter-wave massive MIMO systems,'' \textit{IEEE Trans. Veh. Technol.}, vol. 65, no. 7, pp. 5731-5737, July 2016.

\bibitem{Ayach Glob 13} O. El Ayach, R. W. Heath Jr., S. Rajagopal, and Z. Pi, ``Multimode precoding in millimeter wave MIMO transmitters with multiple antenna sub-arrays,'' in \textit{Proc. IEEE Global Commun. Conf. (GLOBECOM)}, Atlanta, USA, Dec. 2013, pp. 3476-3480.



%
%%%% quantized phase + discrete beam
%
%\bibitem{Sohrabi JSAC 16} F. Sohrabi and W. Yu, ``Hybrid digital and analog beamforming design for large-scale antenna arrays,'' \textit{IEEE J. Sel. Topics in Signal Process.}, vol. 10, no. 3, pp. 501每513, April 2016.
%
%
%\bibitem{Lin TSP} Y.-P. Lin, ``On the quantization of phase shifters for hybrid precoding systems,'' \textit{IEEE Trans. Signal Process.}, to Appeal.
%
%
%\bibitem{Gao TVT 16} X. Gao, L. Dai, C. Yuen, and Z. Wang, ``Turbo-like beamforming based on Tabu search algorithm for millimeter-wave massive
%MIMO systems, \textit{IEEE Trans. Veh. Technol.}, vol. 65, no. 7, pp. 5731-5737, July 2016.
%
%
%
%\bibitem{Kim Globecom 13} T. Kim, J. Park, J.-Y. Seol, S. Jeong, J. Cho, and W. Roh, ``Tens of Gbps support with mmWave beamforming
%systems for next generation communications,'' in \textit{Proc. IEEE Globecom},  Atlanta, GA, Dec. 2013.
%
%
%\bibitem{Wang JSAC 09} J. Wang \textit{et al.}, ``Beam codebook based beamforming protocol for multi-Gbps millimeter-wave WPAN systems,'' \textit{IEEE J. Sel. Areas Commun.}, vol. 27, no. 8, pp. 1390每1399, Oct. 2009.
%
%\bibitem{Cordeiro 10}  C. Cordeiro, D. Akhmetov, and M. Park, ``IEEE 802.11 ad: Introduction and performance evaluation of the first multi-Gbps WiFi technology,'' in \textit{Proc. ACM Int. Workshop mmWave Commun.}, Chicago, IL, Sept. 2010, pp. 3每8.





%%%% sub-array


%
%\bibitem{Zhou VTC 15} L. Zhou and Y. Ohashi, ``Fast codebook-based beamforming training for mmWave MIMO systems with subarray structures,'' in \textit{Proc. IEEE Veh. Technol. Conf. (VTC)}, Boston MA, Sept. 2015.
%

%
%\bibitem{Park arXiv 16} S. Park, A. Alkhateeb, and R. W. Heath Jr., ``Dynamic subarrays for hybrid precoding in wideband mmWave MIMO systems,'' arXiv.
%
%
%\bibitem{Han CL 14} S. Han, C.-L. I, Z. Xu, and S. Wang, ``Reference signals design for hybrid analog and digital beamforming,'' \textit{IEEE Commun. Lett.}, vol. 18, no. 7, pp. 1191-1193, July 2014.
%
%\bibitem{Singh TWC 15} J. Singh and S. Ramakrishna, ``On the feasibility of codebook-based beamforming in millimeter wave systems with multiple antenna arrays,'' \textit{IEEE Trans. Wireless Commun.}, vol. 14, no. 5, pp. 2670每2683, May 2015.
%
%
%\bibitem{Kim Globalcom 13} C. Kim, T. Kim, and J.-Y. Seol, ``Multi-beam transmission diversity with hybrid beamforming for MIMO-OFDM systems,'' in \textit{Proc. IEEE Global Commun. Conf. Workshops (GLOBECOM Wkshps)}, Atlanta, GA, Dec. 2013, pp. 61每65.
%
%
%\bibitem{Han CM 15} S. Han, C.-L. I, Z. Xu, and C. Rowell, ``Large-scale antenna systems with hybrid analog and digital beamforming for millimeter wave 5G,'' \textit{IEEE Commun. Mag.}, vol. 53, no. 1, pp. 186每194, Jan. 2015.
%
%\bibitem{Zhang TCOM 15} J. Zhang, X. Huang, V. Dyadyuk, and Y. Guo, ``Massive hybrid antenna array for millimeter-wave cellular communications,'' \textit{IEEE Wireless Commun.}, vol. 22, no. 1, pp. 79每87, Feb. 2015.
%
%\bibitem{Tamijani TAP 03} A. Abbaspour-Tamijani and K. Sarabandi, ``An affordable millimeterwave beam-steerable antenna using interleaved planar subarrays,'' \textit{IEEE Trans. Antennas Propag.}, vol. 51, no. 9, pp. 2193每2202, Nov. 2003.

\end{thebibliography}
\end{document}